\documentclass[aps,prb,twocolumn,superscriptaddress,amsmath]{revtex4-1}

\usepackage{graphicx}

\begin{document}


\title{Spin injection from a normal metal into a mesoscopic superconductor}



\author{M. J. Wolf}
\affiliation{Institut f\"ur Nanotechnologie, Karlsruher Institut f\"ur Technologie, Karlsruhe, Germany}
\author{F. H\"ubler}
\affiliation{Institut f\"ur Nanotechnologie, Karlsruher Institut f\"ur Technologie, Karlsruhe, Germany}
\affiliation{Center for Functional Nanostructures, Karlsruher Institut f\"ur Technologie, Karlsruhe, Germany}
\affiliation{Institut f\"ur Festk\"orperphysik, Karlsruher Institut f\"ur Technologie, Karlsruhe, Germany}
\author{S. Kolenda}
\affiliation{Institut f\"ur Nanotechnologie, Karlsruher Institut f\"ur Technologie, Karlsruhe, Germany}
\author{H. v. L\"ohneysen}
\affiliation{Center for Functional Nanostructures, Karlsruher Institut f\"ur Technologie, Karlsruhe, Germany}
\affiliation{Institut f\"ur Festk\"orperphysik, Karlsruher Institut f\"ur Technologie, Karlsruhe, Germany}
\affiliation{Physikalisches Institut, Karlsruher Institut f\"ur Technologie, Karlsruhe, Germany}
\author{D. Beckmann}
\email[e-mail address: ]{detlef.beckmann@kit.edu}
\affiliation{Institut f\"ur Nanotechnologie, Karlsruher Institut f\"ur Technologie, Karlsruhe, Germany}
\affiliation{Center for Functional Nanostructures, Karlsruher Institut f\"ur Technologie, Karlsruhe, Germany}

\date{\today}

\begin{abstract}
We report on nonlocal transport in superconductor hybrid structures, with ferromagnetic as well as normal-metal tunnel junctions attached to the superconductor. In the presence of a strong Zeeman splitting of the density of states, both charge and spin imbalance is injected into the superconductor. While previous experiments demonstrated spin injection from ferromagnetic electrodes, we show that spin imbalance is also created for normal-metal injector contacts. Using the combination of ferromagnetic and normal-metal detectors allows us to directly discriminate between charge and spin injection, and demonstrate a complete separation of charge and spin imbalance. The relaxation length of the spin imbalance is of the order of several $\mu$m and is found to increase with a magnetic field, but is independent of temperature. We further discuss possible relaxation mechanisms for the explanation of the spin relaxation length.
\end{abstract}


\maketitle


\section{Introduction}
When a spin-polarized current is injected from a ferromagnet into a spin-degenerate metal, it creates a non-equilibrium spin accumulation which is described by a difference in the occupation probabilities of states for spin up and down. Spin injection from ferromagnets has been demonstrated both for normal metals\cite{johnson1985} and spin-degenerate superconductors.\cite{johnson1994} Recently, spin injection from ferromagnets into superconductors in the presence of a strong Zeeman splitting has been observed.\cite{huebler2012b,quay2012} In this case, spin accumulation mainly originates from the spin-dependent density of states in the superconductor,\cite{giazotto2008} rather than a difference in occupation probabilities. In this article, we extend our previous work and demonstrate efficient spin injection from a normal metal into a superconductor with a large Zeeman splitting. Using both ferromagnetic and normal-metal contacts, we also show the separation of spin and charge imbalance in a single sample.

\section{Samples and Experiment}

\begin{figure}[h]
\includegraphics[width=\columnwidth]{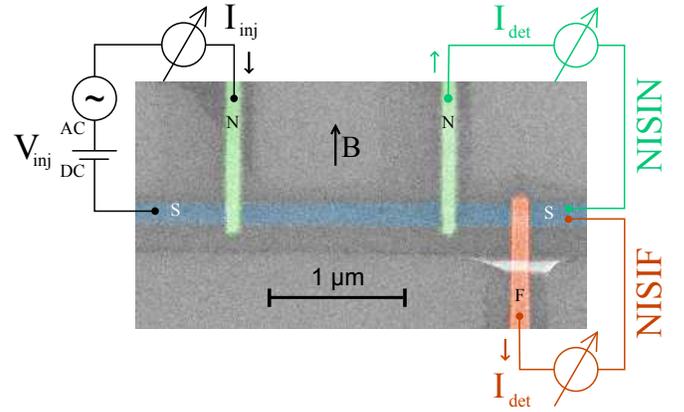}
\caption{\label{fig_sem}(color online) Scanning electron microscopy image of a section of sample A. An iron (F) and two copper (N) electrodes form tunnel contacts with a superconducting aluminum (S) wire. Examples of two measurement configurations for nonlocal measurements using a normal-metal injector and either a normal-metal (NISIN) or ferromagnetic (NISIF) detector are indicated.}
\end{figure}

Samples were fabricated by e-beam lithography and shadow evaporation techniques. In a fist step, a superconducting (S) aluminum wire of thickness $t_\mathrm{Al} \approx 15$~nm  is created. The  aluminum wire is oxydized \textit{in situ} to form a thin but pinhole-free tunnel barrier by exposing it to $0.5$~Pa of pure oxygen for five minutes. After the oxidation, counterelectrodes of ferromagnetic (F) iron and nonmagnetic (N) copper are deposited under a second and third angle respectively. Care was taken to have no overlaps between the N and F electrodes in the proximity of the tunnel contacts. 

We investigated samples with a different number of N and F electrodes and variations in the contact distances, but otherwise similar parameters. An overview of the sample parameters is given in Table~\ref{tab_sample_properties}. Sample C, which has only ferromagnetic junctions, is the same as the sample labeled FISIF in Ref.~\onlinecite{huebler2012b}, and is included here only for comparison with the new results on samples with mixed junction types. Figure~\ref{fig_sem} shows a scanning electron microscopy image of a part of sample A as well as a scheme of the measurement setup. The results presented in this paper stem from sample A unless explicitly stated.

All measurements were performed in a dilution refrigerator at temperatures down to $T = 50$~mK with the magnetic field in the plane of the contacts parallel to the iron leads, as indicated in Fig.~\ref{fig_sem}. A voltage $V_\mathrm{inj}$ consisting of a dc bias and a low-frequency ac excitation was applied to one tunnel contact, called injector, and the ac part of the resulting current $I_\mathrm{inj}$ was measured by standard lock-in techniques to obtain the local conductance $g_\mathrm{loc} = dI_\mathrm{inj} / dV_\mathrm{inj}$. Simultaneously, the ac current $I_\mathrm{det}$ through a second contact, called detector, was measured to determine the nonlocal conductance $g_{\textrm{nl}} = dI_\mathrm{det} / dV_\mathrm{inj}$. For details of the setup see Ref. \onlinecite{brauer2010}. To account for the slight variation in the conductance from contact to contact, we plot the normalized nonlocal conductance $\hat{g}_{\textrm{nl}} = g_{\textrm{nl}} / G_\mathrm{inj} G_\mathrm{det}$ throughout the paper, were $G_\mathrm{inj}$ and $G_\mathrm{det}$ are the normal-state conductances of the two junctions.

The nonlocal conductance was measured for different contact configurations, where both injector and detector could be either normal (N) or ferromagnetic (F). These configurations will be labeled by $A$ISI$A$, where $A$ and $A$ denote the injector and detector contacts, respectively. Two possible configurations with a normal-metal injector and normal-metal (NISIN) or ferromagnetic (NISIF) detector are shown as examples in Fig.~\ref{fig_sem}. The distance between injector and detector contact is denoted by $d$.

\begin{table*}[bt] 
\caption{\label{tab_sample_properties}Overview of sample properties. Junction properties: number of normal metal (N) and ferromagnetic (F) tunnel junctions, range of normal-state tunnel conductances $G_\textrm{N/F}$, and range of contact distances $d$. Aluminum film properties: film thickness $t_\textrm{Al}$, resistivity $\rho_\textrm{Al}$, critical temperature $T_c$, critical magnetic field $B_c$, coherence length $\xi$,  diffusion constant $D$.}
\begin{ruledtabular}
\begin{tabular}{llccccccccc}
        &           & $G_\textrm{F}$ & $G_\textrm{N}$ & $d$        & $t_\textrm{Al}$ & $\rho_\textrm{Al}$ & $T_c$ & $B_c$ & $\xi$ & $D$        \\
Sample  & junctions & ($\mu$S)       & ($\mu$S)       & ($\mu$m)   & (nm)            & ($\mu\Omega$cm)    & (K)   & (T)   & (nm)  & (cm$^2$/s) \\ \hline
A       & 2F, 4N    &  $300-330$     & $340-360$      & $0.5-7.0$  & 12              & 13.5               & 1.6   & 2.1   & 63    & 14.4       \\
B       & 1F, 5N    &  400           & $170-310$      & $0.3-10.3$ & 16.5            & 9.8                & 1.5   & 1.5   & 76    & 19.9       \\
C       & 5F        &  $510-550$     &                & $0.5-8.0$  & 14              & 8.4                & 1.6   & 2.1   & 79    & 23.1       \\
\end{tabular}
\end{ruledtabular}
\end{table*}

\section{Model}
Before showing the experimental results, we would like to briefly describe the model we have used to analyze our data. 

In order to describe the local conductance of the injector junctions, we use the theory of tunneling in superconductors in high magnetic field.\cite{maki1964b,meservey1994} The contribution of a single spin projection $\sigma=\pm 1$ to the tunnel conductance is given by
\begin{equation}
  g_\sigma = \frac{G_\mathrm{inj}}{2}\left(1-\sigma P_\mathrm{inj}\right) \int n_\sigma(E) f' dE,
  \label{eqn_gsigma}
\end{equation}
where $P_\mathrm{inj}$ is the spin polarization of the tunnel conductance, $n_\sigma(E)$ is the normalized quasiparticle density of states in the superconductor for a single spin, and
\begin{equation}
 f'=\frac{\partial\left[-f_0(E+eV_\mathrm{inj})\right]}{\partial eV_\mathrm{inj}}
\end{equation}
is the derivative of the Fermi function. The density of states $n_\sigma(E)$ is calculated by standard methods, including the pair-breaking parameter $\Gamma$, the spin-orbit scattering strength $b_\mathrm{so}$, and the Zeeman effect.\cite{maki1964b,meservey1994} The injector conductance is given by the sum of the two spin contributions, 
\begin{equation}
 g_\mathrm{loc}=g_\downarrow+g_\uparrow,\label{eqn_gloc}
\end{equation}
whereas the differential spin current is proportional to their difference
\begin{equation}
g_\downarrow-g_\uparrow\propto\int \left[ P_\mathrm{inj}\left(n_\downarrow+n_\uparrow\right)+\left(n_\downarrow-n_\uparrow\right)\right] f' dE.\label{eqn_spin_injection}
\end{equation}
As mentioned in the introduction, two terms contribute to spin injection. The the first term is responsible for spin injection from a ferromagnet into a spin-degenerate superconductor.\cite{johnson1994,poli2008,yang2010} The second term is due to the spin-dependence of the density of states. It appears only in the presence of a Zeeman splitting, but does not require a ferromagnet for spin injection. In previous experiments on high field spin-injection into superconductors using ferromagnetic junctions,\cite{huebler2012b,quay2012} both terms contributed to spin injection, while in the present experiment we probe exclusively the second term.

The injected quasiparticles create both charge and spin imbalance in the superconductor. We describe both on an equal footing by a straightforward extension of the simple models discussed in Refs.~\onlinecite{huebler2012b} and \onlinecite{huebler2010}. The densities of non-equilibrium charge and spin for each spin band are denoted by $Q^*_\sigma$ and $S_\sigma$, respectively. Nonequilibrium charge and spin relax over time scales $\tau_{Q^*}$ and $\tau_{S}$, respectively, leading to an exponential decay over the two relaxation lengths $\lambda_{Q^*}=\sqrt{D\tau_{Q^*}}$ and $\lambda_{S}=\sqrt{D\tau_{S}}$, where $D$ is the diffusion constant of the superconductor. For one-dimensional diffusion along the superconducting wire in the geometry of our experiment, their injection rates are given by
\begin{equation}
\frac{d\dot{Q}^*_\sigma}{dV_\mathrm{inj}} = \frac{f^*_\sigma g_\sigma}{2 e \mathcal{A} \lambda_{Q^*}}\,\,\mathrm{and}\,\,\frac{d\dot{S}_\sigma}{dV_\mathrm{inj}} = \frac{g_\sigma}{2 e \mathcal{A} \lambda_{S}} ,
\end{equation}
where $f^*_\sigma$ accounts for the fractional quasiparticle charge,\cite{huebler2010} and $\mathcal{A}$ is the cross section of the superconducting wire. The current flowing out of the detector is given by\cite{zhao1995,huebler2012b}
\begin{equation}
  I_\mathrm{det} = \frac{G_\mathrm{det}}{N_0 e} \left[(  Q^*_\downarrow + Q^*_\uparrow ) + P_\mathrm{det} (S_\downarrow-S_\uparrow) \right].
  \label{eqn_Igeneral}
\end{equation}
Combining injection, relaxation and detection in the same way as in Ref. \onlinecite{huebler2010}, we obtain
\begin{equation}
\hat{g}_{\textrm{nl}} = \frac{1}{G_\mathrm{inj}G_\mathrm{det}}\frac{dI_\mathrm{det}}{dV_\mathrm{inj}}=\hat{g}_{\textrm{nl}}^{Q^*}+\hat{g}_{\textrm{nl}}^{S},\label{eqn_g_nl}
\end{equation}
where the contribution due to charge imbalance is
\begin{equation}
\hat{g}_{\textrm{nl}}^{Q^*} = \frac{f^*_\downarrow g_\downarrow+f^*_\uparrow g_\uparrow}{G_\mathrm{inj}} \frac{\rho_\textrm{N} \lambda_{Q^*}}{2 \mathcal{A}} \exp(-d/\lambda_{Q^*})\label{eqn_g_nl_calc_CI}
\end{equation}
and the contribution due to spin imbalance is
\begin{equation}
\hat{g}_{\textrm{nl}}^{S}= \frac{g_\downarrow-g_\uparrow}{G_\mathrm{inj}} P_\mathrm{det} \frac{\rho_\textrm{N} \lambda_{S}}{2 \mathcal{A}} \exp(-d/\lambda_{S}).\label{eqn_g_nl_calc_SI}
\end{equation}
Here $d$ is the distance between the contacts, and $\rho_\textrm{N}$ is the normal-state resistivity of the superconductor.
These two equations form the basis of our data analysis. We note that the charge-imbalance signal is always positive, whereas the spin-imbalance signal can have either sign. Also, detection of the spin signal requires a finite $P_\mathrm{det}$. Therefore, a normal-metal detector will measure only the charge signal, whereas a ferromagnetic detector measures the sum of the charge and spin signals. Comparing the signals of ferromagnetic and nonmagnetic junctions we can therefore discriminate the two contributions.

\section{Results}

\begin{figure}[h]
\includegraphics[width=\columnwidth]{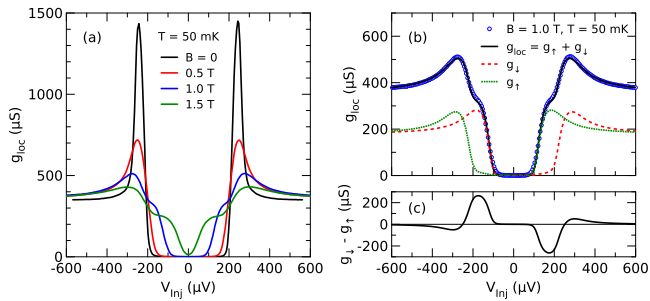}
\caption{\label{fig_loc}
(color online) 
(a) Differential conductance $g_\mathrm{loc}$ of an NIS junction as a function of the bias voltage $V_\mathrm{inj}$ for different magnetic fields $B$.
(b) Data for $B= 1.0$~T together with a fit to Eq. (\ref{eqn_gloc}) and the contributions of the individual spin orientations.
(c) Calculated spin injection $g_\downarrow - g_\uparrow$.}
\end{figure}

Figure \ref{fig_loc}(a) shows the differential conductance of an NIS tunnel junction for different applied magnetic fields $B$. At zero field, $g_\textrm{loc}$ is negligible for injector bias in the subgap region, $|V|\lesssim 200~\mathrm{\mu V}$, demonstrating the high quality of the pinhole-free tunnel barrier. Above the gap, sharp singularities are seen before the conductance drops back to its normal state value at high bias. In an applied magnetic field, the density of states broadens due to pair-breaking, and the Zeeman splitting is clearly seen for large fields.\cite{meservey1975} The data for $B=1.0$~T are given in Fig.~\ref{fig_loc}(b) together with a fit to Eq. (\ref{eqn_gloc}). Details of the fitting procedure are given in Refs. \onlinecite{brauer2010,huebler2012}.
From the fit, the individual contributions $g_\uparrow$ and $g_\downarrow$ to the tunnel conductance are extracted and indicated in Fig.~\ref{fig_loc}(b) as dotted and dashed curves, respectively. The difference $g_\downarrow - g_\uparrow$ is plotted in Fig.~\ref{fig_loc}(c). In the bias window of the Zeeman splitting around $V_\textrm{inj} \approx \pm 180~\mathrm{\mu V}$, a single spin band dominates conductance, and consequently the spin injection has a maximum. $g_\downarrow-g_\uparrow$ is the only bias-dependent quantity which enters Eq. (\ref{eqn_g_nl_calc_SI}), thus the nonlocal signal due to spin imbalance is expected to have the same shape as depicted here. 

\begin{figure}[h]
\includegraphics[width=\columnwidth]{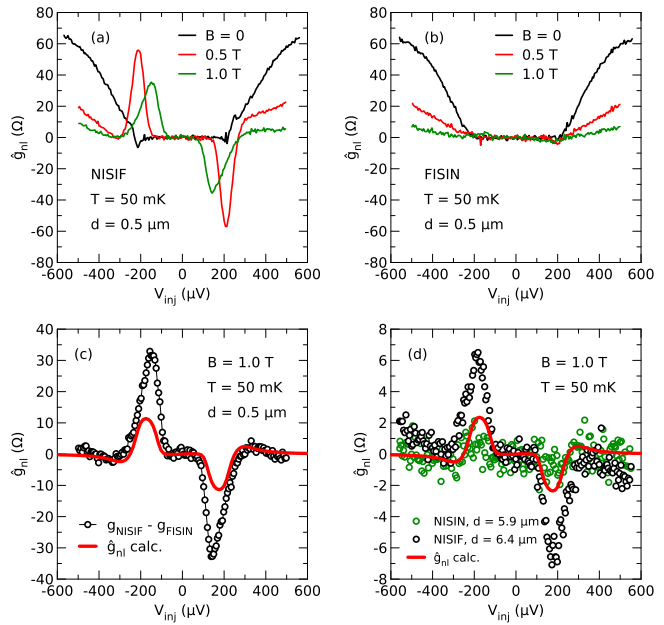}
\caption{\label{fig_NISIF_FISIN} (color online) 
(a) Normalized nonlocal differential conductance $\hat{g}_{\textrm{nl}}$ for  the closest pair of electrodes with N as injector, F as detector (NISIF) and for several magnetic fields. 
(b) $\hat{g}_{\textrm{nl}}$ for the same contacts and magnetic fields, but now with F as injector and N as detector (FISIN).
(c) Difference of the signals from (a) and (b) at $B=1.0~\mathrm{T}$.
(d) Comparison of NISIF and NISIN data at $B = 0.5$ T for large contact separation $d$. The lines in (c) and (d) are model predictions (see discussion).}
\end{figure}

Figure~\ref{fig_NISIF_FISIN}(a) and (b) shows the nonlocal conductance for the combination of a ferromagnetic and a normal-metal contact (the two right-most contacts in Fig. \ref{fig_sem}). In panel (a), the normal-metal contact was used as injector, and the ferromagnetic contact was used as detector (NISIF), whereas in panel (b) their roles are reversed (FISIN). For $B = 0$, the nonlocal signal shown in Fig.~\ref{fig_NISIF_FISIN}(a) is zero for bias values below the energy gap of the superconductor and increases steeply as soon as the injector bias exceeds the energy gap. In an applied magnetic field, additional asymmetric peaks appear at voltages near the gap, which first grow in height until $B \approx 0.5$~T. For higher fields, they start to decline, broaden and move inwards.
If one swaps the role of injector and detector, Fig.~\ref{fig_NISIF_FISIN}(b), only the symmetric part of the signal is found, which quickly decreases with applied magnetic field. This signal is due to charge imbalance, as has been shown in previous experiments on NISIN structures.\cite{huebler2010,huebler2012b} Using Eq. (\ref{eqn_g_nl}) we can extract the spin signal alone by subtracting the charge signal seen in Fig.~\ref{fig_NISIF_FISIN}(b) from the charge+spin signal in Fig.~\ref{fig_NISIF_FISIN}(a). This is highlighted by Fig.~\ref{fig_NISIF_FISIN}(c) for the data at $B = 1.0$~T. The difference signal shows the asymmetric peak structure as well as slight side extrema of different sign than the peaks. 

A direct comparison of the charge and spin signals for large contact separations $d\approx 6~\mathrm{\mu m}$ is given in Fig.~\ref{fig_NISIF_FISIN}(d). Here, the two right-most contacts in Fig. \ref{fig_sem} were used as detectors, and the same normal-metal contact was used for injection. Therefore, the nonequilibrium quasiparticle populations probed by the two detectors are essentially the same. For these contact distances $d$, the charge imbalance has relaxed and the NISIN signal has thus vanished. In contrast, the asymmetric peak structure of the NISIF signal is still visible. This directly probes the spatial separation of charge and spin imbalance. 

The solid curves in Fig.~\ref{fig_NISIF_FISIN}(c) and (d) are calculations of the nonlocal signal according to Eq. (\ref{eqn_g_nl_calc_SI}) and will be discussed below.

\begin{figure}[hbt]
\includegraphics[width=\columnwidth]{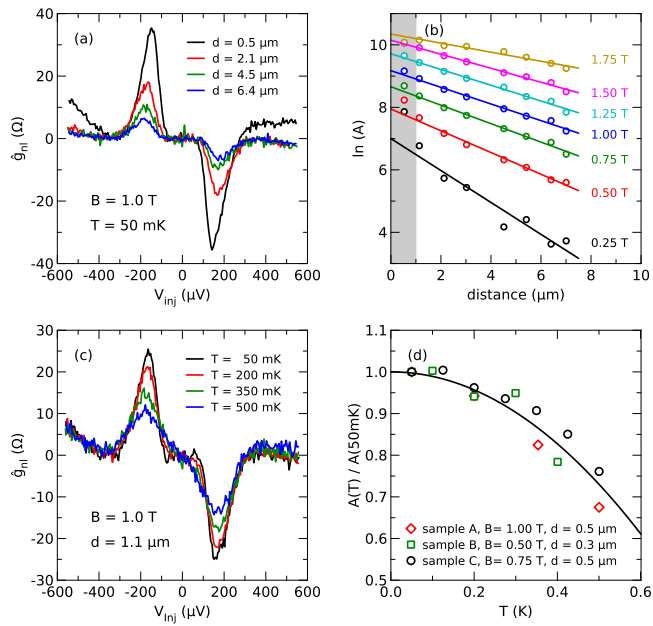}
\caption{\label{fig_dist+T} (color online)
(a) Nonlocal conductance $\hat{g}_{\textrm{nl}}$ for several contact distances $d$ at fixed magnetic field $B = 1.0$~T. 
(b) Semilogarithmic plot of the peak area $A$ as a function of contact distance $d$ for different magnetic fields $B$. The lines are fits to an exponential decay.
(c) Nonlocal conductance $\hat{g}_{\textrm{nl}}$ for different temperatures at fixed magnetic field and contact distance. 
(d) Normalized peak area $A(T)/A(T=50~\mathrm{mK})$ for small contact distances $d$ as a function of temperature $T$, shown for all three samples. The line is a guide to the eye.}
\end{figure}

In Fig.~\ref{fig_dist+T}(a) the nonlocal conductance  $\hat{g}_{\textrm{nl}}$  is shown for different contact distances $d$ at a fixed magnetic field $B = 1.0$~T. The symmetric high-bias signal due to charge imbalance is visible only for the shortest injector-detector distance of $d = 0.5\,\mu$m whereas the asymmetric peaks are present at the same injector bias value for all distances and are visible even for the largest distance of $d = 6.4\,\mu$m. 

To further analyze the dependence of the signal on contact distance, we extracted the charge signal by calculating the symmetric part $\hat{g}_\textrm{s}=\left[\hat{g}_\textrm{nl}(V_\textrm{inj}) + \hat{g}_\textrm{nl}(-V_\textrm{inj})\right]/2$, and the spin signal by calculating the antisymmetric part $\hat{g}_\textrm{a}=\left[\hat{g}_\textrm{nl}(V_\textrm{inj}) - \hat{g}_\textrm{nl}(-V_\textrm{inj})\right]/2$. For the spin signal, we then calculated the peak area $A$ by integrating   $\hat{g}_{\textrm{a}}$ numerically from  $V_\textrm{inj}=-290\,\mu$V to $V_\textrm{inj}=0\,\mu$V. 

\begin{figure}[hbt]
\includegraphics[width=\columnwidth]{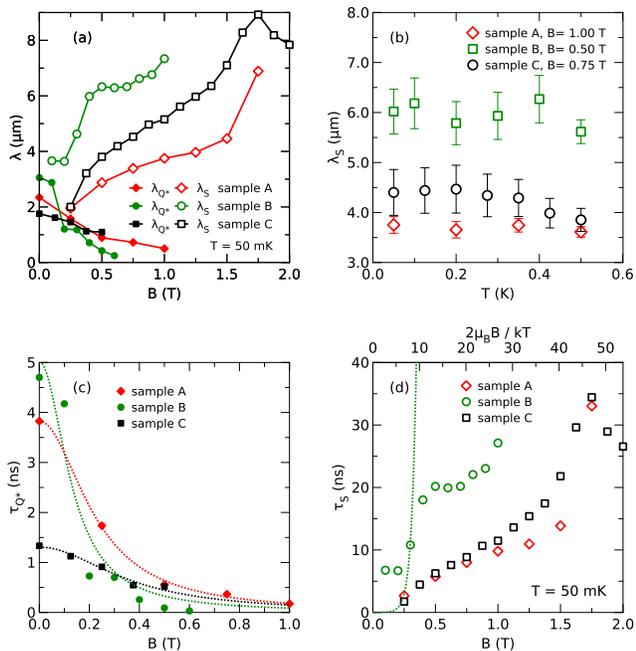}
\caption{\label{fig_lengths} (color online)
(a) Spin (open symbols) and charge imbalance (closed symbols) relaxation lengths $\lambda_{Q^*}$ and $\lambda_{S}$ as a function of the magnetic field $B$.
(b) Spin relaxation length $\lambda_{S}$ as a function of temperature $T$ at fixed magnetic field $B$.
(c) Charge imbalance relaxation time $\tau_{Q^*}$ as a function of the magnetic field $B$, together with fits (see text).
(d) Spin relaxation time $\tau_{S}$ as a function of the magnetic field $B$, with an attempt to fit with $\exp(2\mu_\textrm{B}B/k_\textrm{B}T)$.}
\end{figure}

Figure~\ref{fig_dist+T}(b) shows the peak area $A$ as a function of contact distance $d$ for different magnetic fields $B$ on a semilogarithmic scale. For better visibility, the datasets have been offset vertically. The data can be fit to an exponential decay, shown by the lines in Fig.~\ref{fig_dist+T}(b), except for the closest contact distance, which was therefore excluded from the fit. From these fits the spin-imbalance relaxation length $\lambda_{S}$ is extracted. To extract the charge-imbalance relaxation length $\lambda_{Q^*}$, we also fit $\hat{g}_\textrm{s}$ at $V_\textrm{inj} = 520~\mu$V to an exponential decay (not shown). The results of these fits are shown in Fig.~\ref{fig_lengths}. 


In Fig.~\ref{fig_dist+T}(c) we show the nonlocal conductance $\hat{g}_{\textrm{nl}}$ for a fixed magnetic field of $B = 1.0$~T and different temperatures $T$. The influence of temperature on the spin signal is to decrease the peak height and broaden the peaks. However, the influence is not just a thermal broadening, as the peak area decreases significantly when the temperature is increased. To analyze the decrease in greater detail, we normalized the peak area $A$ to the value at $T = 50$~mK and found a similar decrease with temperature for all three samples as can be seen in Fig.~\ref{fig_dist+T}(d). 

The relaxation lengths $\lambda_{S}$ and $\lambda_{Q^*}$ obtained from the exponential fits described above are shown in Fig.~\ref{fig_lengths}(a) as a function of magnetic field $B$ for all three samples. $\lambda_{S}$ increases monotonously with $B$, but one observes for all three samples a change in slope at $B \approx 0.5$~T, which is most pronounced for sample B.
For the highest fields close to $B_c$, the behavior gets less systematic. Since the absolute values of the peak areas decrease significantly when approaching $B_c$, the reduced signal-to-noise ratio serves as a plausible explanation. $\lambda_{Q^*}$, in contrast, is largest for $B = 0$ and reduces quickly in a magnetic field. While $\lambda_S$  show a pronounced dependence on the magnetic field $B$, no significant change with temperature $T$ can be found, as it is shown for all three samples in Fig.~\ref{fig_lengths}(b). 

Figures~\ref{fig_lengths}(c) and (d) present the charge-imbalance and spin-imbalance relaxation times $\tau_{Q^*}$ and $\tau_{S}$. These were determined from the relaxation lengths shown in  Fig.~\ref{fig_lengths}(a) using the relations $\tau_{Q^*}=\lambda^2_{Q^*}/D$ and $\tau_{S}=\lambda^2_{S}/D$ and the known diffusion coefficients $D$ given in Table~\ref{tab_sample_properties}.  

At zero magnetic field, the charge imbalance relaxation rate $\tau_{Q^*}^{-1}$ is determined by the combined effect of inelastic scattering and elastic impurity scattering in conjunction with the gap anisotropy of aluminum.\cite{chi1979} In addition, magnetic pair-breaking perturbations increase the relaxation rate.\cite{schmid1975} Previous experiments on charge imbalance have shown that these two contributions are additive at low temperature, i.e., $\tau_{Q^*}^{-1}=\tau_0^{-1}+\beta\Gamma$, where $\tau_0^{-1}$ denotes the relaxation rate at zero field, $\Gamma$ is the pair breaking parameter, and $\beta$ is a coherence factor.\cite{huebler2010} Using $\Gamma = (B/B_c)^2/2$ for a thin film in parallel magnetic field, we obtain $\tau_{Q^*} = (\tau_0^{-1}+bB^2)^{-1}$. Fits to this dependence are shown in Fig.~\ref{fig_lengths}(c). As can be seen, $\tau_{Q^*}$ is a few nanoseconds at zero field, and then rapidly drops at higher fields.

Figures~\ref{fig_lengths}(d) shows the spin relaxation time $\tau_{S}$ as a function of magnetic field $B$, where we also show the ratio of Zeeman splitting and temperature, $2\mu_\textrm{B}B/k_\textrm{B}T$, at the top of the panel. $\tau_{S}$ is a few nanoseconds at small fields, similar to $\tau_{Q^*}$. At higher fields, it increases to $10-20~\mathrm{ns}$, and exceeds  $\tau_{Q^*}$ by at least two orders of magnitude. When the Zeeman splitting is not much larger than the thermal broadening, we expect $S_\downarrow/S_\uparrow\propto\exp(2\mu_\textrm{B}B/k_\textrm{B}T)$. Therefore, the relaxation time might show a similar dependence on magnetic field, as assumed in Ref.~\onlinecite{quay2012}. To check this assumption, we have plotted an attempt to fit $\tau_{S}\propto\exp(2\mu_\textrm{B}B/k_\textrm{B}T)$ in Fig.~\ref{fig_lengths}(d). As can be seen, the fit is possible at best in small fields, but clearly fails to describe the high-field data.

\section{Discussion}

In our previous work,\cite{huebler2012b} we reported on spin injection from ferromagnetic contacts into a superconductor in the presence of a Zeeman splitting. In that case, both the finite spin polarization of the injector, $P_\mathrm{inj}$, and the spin-dependence of the density of states, $n_\downarrow-n_\uparrow$, contribute to spin injection, as inferred from Eq. (\ref{eqn_spin_injection}). $n_\downarrow-n_\uparrow$ is responsible for the asymmetric shape of the signals. As explained in Ref.~\onlinecite{huebler2012b}, the role of $P_\mathrm{inj}$ is only to increase the amplitude of the positive peak, and to decrease the amplitude of the negative peak (in other words, there is an overall upward shift of the signal). As can be seen in Figs.~\ref{fig_NISIF_FISIN} and \ref{fig_dist+T}, the amplitudes of the  peaks at positive and negative bias are the same in the present experiment, where spin is injected from a normal metal with $P_\mathrm{inj}=0$.

For a quantitative comparison with the model, the nonlocal signals calculated from Eq. (\ref{eqn_g_nl_calc_SI}) for a spin-degenerate injector are plotted as solid lines in Figs.~\ref{fig_NISIF_FISIN}(c) and (d). The factors $(g_\downarrow-g_\uparrow)/G_\mathrm{inj}$ and $P_\textrm{det}$ are obtained from the fits of the local conductance of the injector and detector junctions, as shown for example in Fig.~\ref{fig_loc}(b). The relaxation length $\lambda_{S}$ is obtained from the exponential fits in Fig.~\ref{fig_dist+T}(b). The normal-state resistance per length of the aluminum wire is known from measurements at $T=4.2~\mathrm{K}$, so that we can calculate the factor $\rho_\textrm{N} \lambda_{S}/2\mathcal{A}$. Thus the nonlocal signal $\hat{g}_{\textrm{nl}}^{S}$ predicted by our simple model can be calculated without free fitting parameters. The shapes of the calculated and measured signals agree qualitatively, whereas the amplitude of the calculated signal is too small. Since the spin injection and detection factors as well as the normal-state properties are known quite accurately, we suspect that the assumption of a single exponential decay length $\lambda_{S}$ independent of energy and bias conditions is the most likely culprit for the disagreement. To elucidate this further, we would like to discuss some possible relaxation mechanisms of the spin signal.

\begin{figure}[hbt]
\includegraphics[width=\columnwidth]{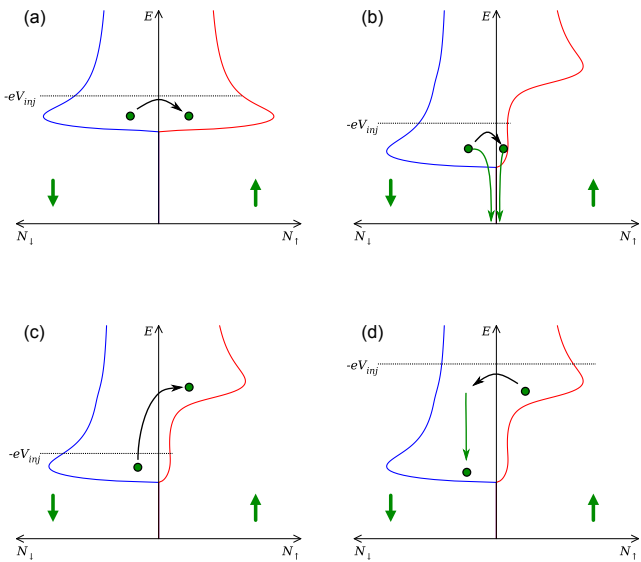}
\caption{\label{fig_rel} (color online) Schematic view of the spin-resolved density of states, including Zeeman splitting and spin-orbit scattering, and various possible relaxation mechanisms. States are partially occupied up to $E=-eV_\mathrm{inj}$  (see discussion for details).
(a) Spin-flip scattering in a spin-degenerate superconductor.
(b) Proposed two-stage relaxation mechanism involving spin flips and recombination for bias voltages in the range of the Zeeman splitting.
(c) Inelastic spin flips to the upper Zeeman band.
(d) Two-stage spin-flip and energy relaxation process at high bias.
}
\end{figure}

Nonequilibrium quasiparticles in superconductors are subject to several different scattering processes. Electron-phonon scattering leads to energy relaxation, charge-imbalance relaxation, as well as recombination of quasiparticles to Cooper pairs.\cite{kaplan1976} Spin-orbit scattering leads to elastic spin flips, and is therefore expected to relax spin imbalance.\cite{yafet1983,*yamashita2002,*morten2004,*morten2005} It also modifies the density of states in the presence of Zeeman splitting.\cite{maki1964b} Magnetic pair-breaking perturbations lead to additional charge-imbalance relaxation,\cite{schmid1975} and may affect all other scattering mechanism by changing the density of states and coherence factors. Magnetic impurities in particular also lead to spin flips.\cite{yafet1983,*yamashita2002,*morten2004,*morten2005} The spin accumulation in a single spin band, $S_\sigma$, is proportional to the product of the density of states $n_\sigma$ and the occupation probability $f_\sigma$. Therefore, for the net spin accumulation we have $S_\downarrow-S_\uparrow\propto n_\downarrow f_\downarrow-n_\uparrow f_\uparrow$. 

Without Zeeman splitting, we have $n_\downarrow=n_\uparrow$, and spin accumulation is due to the difference in occupation probability $f_\downarrow-f_\uparrow$. Several experiments have probed spin injection from ferromagnets into spin-degenerate superconductors.\cite{johnson1994,poli2008,yang2010} In this regime, spin relaxation is due elastic spin flips by spin-orbit scattering or magnetic-impurity scattering,\cite{yafet1983,*yamashita2002,*morten2004,*morten2005} as sketched in Fig.~\ref{fig_rel}(a). The spin-relaxation length in the superconducting state is expected to be either the same (spin-orbit scattering) or shorter (magnetic-impurity scattering) than in the normal-state. We have determined the spin-relaxation length in the normal state, $\lambda_\mathrm{N}$, by nonlocal spin-valve experiments for sample C, and found $\lambda_\mathrm{N}=370~\mathrm{nm}$.\cite{huebler2012b} For samples A and B, we could not determine $\lambda_\mathrm{N}$, since these samples have only one or two ferromagnetic junctions. However, the aluminum film parameters are similar, and we can assume $\lambda_\mathrm{N}\sim 400~\mathrm{nm}$ for all three samples.

For spin injection from a normal metal in the presence of a Zeeman splitting, we expect $f_\downarrow=f_\uparrow$, and spin accumulation is solely caused by the difference in density of states, $n_\downarrow-n_\uparrow$. By detailed balance, the net spin-flip rate due to either spin-orbit scattering or magnetic impurity scattering should therefore be zero, and not lead to any spin relaxation at all. Neither energy nor charge relaxation have any effect on spin accumulation. Recombination reduces the overall number of quasiparticles, but does not lead to spin relaxation, as it removes one quasiparticle from each spin band. In the energy window of the Zeeman splitting, however, recombination will deplete the spin-up band much faster than the spin-down band due to the different density of states. Therefore, it will indirectly enable spin flips. One possible spin relaxation mechanism in our experiment is therefore a two-stage process of recombination and spin-flip scattering, as shown schematically in Fig.~\ref{fig_rel}(b). Since the normal-state spin-diffusion length $\lambda_\mathrm{N}$ is much shorter than the observed $\lambda_{S}$, we can assume that recombination is the bottleneck for this mechanism. The impact of spin-orbit scattering on recombination in high fields was considered theoretically for SIS tunnel junctions.\cite{grimaldi1996,*grimaldi1997}

The two stage relaxation mechanism can explain why $\lambda_{S}$ is very large, since recombination is expected to be very slow at low temperatures. However, once $\mu_\textrm{B}B\gg k_\textrm{B}T$, a further increase of the field will essentially no longer change the density of states in the energy range $E<-eV_\mathrm{inj}$ occupied by quasiparticles. Therefore, it is not clear why $\lambda_{S}$ should continue to increase at high magnetic fields. One possible explanation would be inelastic spin flips\cite{fabian1999} to the upper Zeeman band, as shown in Fig.~\ref{fig_rel}(c). The inelastic spin-flip rate should decrease upon increasing field, since larger energy transfer is needed. As second possible explanation is the phonon bottleneck of recombination:\cite{rothwarf1967} For higher fields, the overall number of nonequilibrium quasiparticles in the energy window of the Zeeman splitting increases, and within the two-stage relaxation mechanism sketched above, more recombination phonons would be created. These would in turn reduce the net recombination rate.

Also, as can be seen in Fig.~\ref{fig_NISIF_FISIN}(d), the measured spin signal extends to higher bias than expected from the spin-injection factor $g_\downarrow-g_\uparrow$ obtained from fitting the injector conductance spectra. For bias beyond the energy window of the Zeeman splitting, an almost equal number of quasiparticles with both spins is injected. Cooling due to inelastic scattering will then be more efficient for the spin-down quasiparticles, since a larger density of states is available at low energy. This will free spin-down states at high energy, and might therefore lead to a net spin-flip scattering rate from the spin-up to the spin-down band. This would have the counterintuitive effect that spin-flip scattering can lead to an increase of spin accumulation, as sketched in Fig.~\ref{fig_rel}(d), and therefore explain that the spin signal extends to higher bias than expected.

Finally, we would like to address the temperature dependence. As shown in Fig.~\ref{fig_lengths}(d), the relaxation length $\lambda_S$ is almost independent of temperature from $50~\mathrm{mK}$ to $500~\mathrm{mK}$. In our experiment, we have both $eV_\mathrm{inj}\gg k_\textrm{B}T$ and $\Delta\gg k_\textrm{B}T$. Therefore, the nonequilibrium quasiparticle distribution is determined mostly by bias, and is almost independent of temperature. In addition $\Delta$ is almost constant in this temperature interval, so that also the density of states and coherence factors are almost constant. This may explain the temperature-independence of $\lambda_S$. A similar behavior was found for $\lambda_{Q^*}$.\cite{huebler2010} In contrast to $\lambda_S$, the peak area of the spin signal shown in Fig.~\ref{fig_dist+T}(d) decreases by about 30~\% upon increasing the temperature from $50~\mathrm{mK}$ to $500~\mathrm{mK}$. This can not be accounted for by a change in the injection factor  $g_\downarrow-g_\uparrow$, since thermal broadening will not lead to a change in peak area.

It is obvious from this discussion that spin relaxation in superconductors in high magnetic fields is a complex process, and a detailed quantitative model is beyond the scope of this article. 

\section{Conclusion}

We have shown spin injection and transport in mesoscopic superconductors in the regime of large Zeeman splitting, and investigated in detail the role of spin-polarized and spin-degenerate injector and detector junctions. We have found that spin injection is possible from a normal metal, whereas a ferromagnet is needed as detector to observe spin accumulation.  For spin injection from a normal metal, spin accumulation is purely due to the spin-dependent density of states in the superconductor in high magnetic fields.  Comparing the nonlocal conductance probed by spin-degenerate and spin-polarized detectors, we were able to directly discriminate charge and spin imbalance. The spin relaxation length increases strongly in a magnetic field, but is found to be almost independent of temperature. A detailed explanation of the relaxation mechanisms remains an open question to theory. 

This work was partially supported by the Deutsche Forschungsgemeinschaft under grant BE-4422/1-1.

\bibliography{lit.bib}

\end{document}